\documentclass[twocolumn]{aastex631}

\newcommand{\CIV}{C\,{\small IV}}
\newcommand{\HeII}{He\,{\small II}}
\newcommand{\NV}{N\,{\small V}}
\newcommand{\CII}{C\,{\small II}}
\newcommand{\SiIV}{Si\,{\small IV}}
\newcommand{\NIV}{N\,{\small IV}}
\newcommand{\OIII}{O\,{\small III}}
\newcommand{\NIII}{N\,{\small III}}
\usepackage[T1]{fontenc}
\usepackage{ae,aecompl}
\usepackage{graphicx}
\usepackage{ulem}
\usepackage{chemformula}
\newcommand{\kmsMp}{km\,s$^{-1}$\,Mpc$^{-1}$}

\shorttitle{Jet-brightening in a Merger of Disk Galaxies}
\shortauthors{Lebowitz et al.}
\graphicspath{{./}{figures/}}

\begin{document}

\title{The Dragonfly Galaxy. III. Jet-brightening of a High-redshift Radio Source Caught in a Violent Merger of Disk Galaxies}

\author{Sophie Lebowitz}
\affiliation{Steward Observatory, Department of Astronomy, University of Arizona, 993 North Cherry Avenue, Tucson, AZ, 85721, USA}
\affiliation{Department of Astronomy, The Ohio State University, Columbus OH 43210 USA}

\author{Bjorn Emonts}
\affiliation{National Radio Astronomy Observatory, 520 Edgemont Road, Charlottesville, VA 22903, USA}

\author{Donald M. Terndrup}
\affiliation{Department of Astronomy, The Ohio State University, Columbus OH 43210 USA}

\author{Joseph N. Burchett}
\affiliation{Department of Astronomy, New Mexico State University, Las Cruces, NM 88003, USA}

\author{J. Xavier Prochaska}
\affiliation{University of California, 1156 High Street, Santa Cruz, CA 95064, USA}
\affiliation{Kavli Institute for the Physics and Mathematics of the Universe,
The University of Tokyo, 5-1-5 Kashiwanoha, Kashiwa, 277-8583, Japan}
\affiliation{Division of Science, National Astronomical Observatory of Japan,
2-21-1 Osawa, Mitaka, Tokyo 181-8588, Japan}

\author{Guillaume Drouart}
\affiliation{International Centre for Radio Astronomy Research, Curtin University, 1 Turner Avenue, Bentley, Western Australia 6102, Australia}

\author{Montserrat Villar-Mart\'{i}n}
\affiliation{Centro de Astrobiolog\'{i}a, CSIC-INTA, Ctra. de Torrej\'{o}n a Ajalvir, km 4, 28850 Torrej\'{o}n de Ardoz, Madrid, Spain}

\author{Matthew Lehnert}
\affiliation{Universit\'{e} Lyon 1, ENS de Lyon, CNRS UMR5574, Centre de Recherche Astrophysique de Lyon, F-69230 Saint-Genis-Laval, France}

\author{Carlos De Breuck}
\affiliation{European Southern Observatory, Karl Schwarzschild Strasse 2, 85748 Garching bei M$\ddot{u}$nchen Germany}

\author{Jo\"{e}l Vernet}
\affiliation{European Southern Observatory, Karl Schwarzschild Strasse 2, 85748 Garching bei M$\ddot{u}$nchen Germany}

\author{Katherine Alatalo}
\affiliation{Space Telescope Science Institute, 3700 San Martin Dr, Baltimore, MD 21218, USA}
\affiliation{Johns Hopkins University, Department of Physics and Astronomy, Baltimore, MD 21218, USA}



\begin{abstract}
The Dragonfly Galaxy (MRC\,0152-209), the most infrared-luminous radio galaxy at redshift $z$\,$\sim$\,2, is a merger system containing a powerful radio source and large displacements of gas. We present kpc-resolution data from ALMA and the VLA of carbon monoxide (6-5), dust, and synchrotron continuum, combined with Keck integral-field spectroscopy. We find that the Dragonfly consists of two galaxies with rotating disks that are in the early phase of merging. The radio jet originates from the northern galaxy and brightens when it hits the disk of the southern galaxy. The Dragonfly Galaxy therefore likely appears as a powerful radio galaxy because its flux is boosted into the regime of high-z radio galaxies by the jet-disk interaction. We also find a molecular outflow of $(1100\,\pm\,550)$ M$_{\odot}$\,yr$^{-1}$ associated with the radio host galaxy, but not with the radio hot-spot or southern galaxy, which is the galaxy that hosts the bulk of the star formation. Gravitational effects of the merger drive a slower and longer lived mass displacement at a rate of $(170\,\pm\,40)$ M$_{\odot}$\,yr$^{-1}$, but this tidal debris contain at least as much molecular gas mass as the much faster outflow, namely $M_{\rm H2} = (3\,\pm\,1) \times 10^9\ (\alpha_{\rm CO}/0.8)\ {\rm M}_\odot$. This suggests that both the AGN-driven outflow and mass transfer due to tidal effects are important in the evolution of the Dragonfly system. The Keck data show Ly$\alpha$ emission spread across 100 kpc, and \CIV\ and \HeII\ emission across 35 kpc, confirming the presence of a metal-rich and extended circumgalactic medium previously detected in CO(1-0). 
\end{abstract}

\keywords{High-redshift galaxies -- Galaxy mergers -- Galaxy interactions -- Radio galaxies -- Radio jets -- Radio loud quasars -- Starburst galaxies -- Ultraluminous infrared galaxies -- Circumgalactic medium -- Radio astronomy -- Submillimeter astronomy}


\section{Introduction}

    For decades, high-redshift radio galaxies have served as excellent laboratories for studying the early formation and evolution of galaxies \citep{miley08}. High-redshift radio galaxies, or HzRGs in short, are often defined to have a redshift of $z$\,$\ga$\,2 and a rest-frame radio power of $P_{\rm 500\,MHz\,(rest)} > 10^{27.5}\,{\rm W}\,{\rm Hz}^{-1}$ \citep{miley08}. The bright synchrotron emission from their powerful, steep-spectrum radio sources has long been used as a beacon for tracing the faint optical signatures of massive galaxies and proto-clusters \citep[e.g.,][see also reviews by \citealt{mccarthy93} and \citealt{miley08}]{rottgering94,chambers96,carilli97,pentericci00}. High redshift radio galaxies occupy the high end of galaxy masses, with $M_{*}$\,=\,10$^{11-12}$\,M$_{\odot}$ (\citealt{seymour07}; see also \citealt{pentericci01}, \citealt{debreuck10}, \citealt{rocca13}). They also possess characteristics that can give valuable insight into the process of galaxy evolution, such as high rates of star formation and strong Active Galactic Nucleus (AGN) activity \citep[e.g.,][]{barthel12,drouart16,wilkes13}, as well as jet-driven gas outflows \citep[e.g.,][]{villar99,nesvadba17}. High-$z$ radio galaxies are often found in overdense regions, as expected from the progenitors of giant elliptical galaxies that occupy the centers of galaxy clusters \citep[e.g.,][]{pentericci97,venemans07,hatch09,galametz12,wylezalek13,dannerbauer14}. These properties had indicated that high-$z$ radio galaxies represent the most active episodes in the early evolution of massive galaxies, but more recent studies reveal that many high-$z$ radio galaxies may be on the way to quenching \citep[e.g.,][]{man19,falkendal19}. Accurate techniques for determining the complex spectral energy distributions \cite[e.g.,][]{drouart18} are starting to reveal that many of the radio host galaxies may even fall below the main sequence of star-forming galaxies \citep{falkendal19}.
		
The Dragonfly Galaxy, MRC\,0152-209, is a radio galaxy at $z = 1.92$ that shows extreme characteristics, even for a high-$z$ radio galaxy. Its starburst infrared luminosity is in the regime of Hyper-Luminous Infra-Red Galaxies (HyLIRGs; $L_{\rm IR} \ge 10^{13}\,{\rm L}_{\odot}$), which is roughly an order of magnitude higher than other HzRGs at $z \sim 2$ \citep{drouart14,falkendal19}. This reflects a star formation rate ($SFR$) $\sim 2000\,{\rm M}_{\odot}\,{\rm yr}^{-1}$, after correcting for infrared emission from dust heated by the AGN \citep{drouart14,falkendal19}. Imaging from the Hubble Space Telescope (HST) Near Infrared Camera and Multi-Object Spectrometer (NICMOS) showed that the Dragonfly appears to be a merger system \citep{pentericci01}. The radio source in the Dragonfly Galaxy is $\sim$10\,kpc in size, and only slightly resolved in existing radio images at 4.5 and 8.2 GHz \citep{pentericci00}.
		
The Dragonfly Galaxy also contains a large mass of cold molecular gas, $M_{\rm H_2} \approx 5 \times 10^{10}\ (\alpha_{\rm CO} /0.8)\ {\rm M}_{\odot}$ \citep{emonts11}, with $\alpha_{\rm CO}$\,=$M_{\rm H2}$/$L^{\prime}_{CO}$\,=\,\,0.8 the CO-to-H$_{2}$ conversion factor found for ultra-luminous infrared galaxies (ULIRGS; \citealt{downes98}. This is derived from a CO(1-0) luminosity that is at the high end of what is found in high-$z$ radio galaxies \citep{emonts14}. The CO(1-0) emission revealed a molecular gas reservoir that is spread across $\sim$60\,kpc, likely reflecting widespread tidal debris of cold gas (\citealt{emonts15a}, hereafter Paper {\sc I}). Observations of CO(6-5) with the Atacama Large Millimeter/submillimeter Array (ALMA) in Cycle 2 showed that the Dragonfly Galaxy is a system of three merging galaxies with one containing a powerful radio source (\citealt{emonts15b}, hereafter Paper {\sc II}). This work revealed that a high mass of cold molecular gas is being displaced between two of the merging galaxies at a rate between  1200 and 3000 M$_\odot\ {\rm yr}^{-1}$. This matches the star-formation rate estimated for the system. However, these Cycle 2 data could not distinguish whether the gas kinematics in the Dragonfly Galaxy were caused by gaseous outflows, or by gravitational interactions between two rotating disk galaxies (Paper {\sc II}).

    In this paper, we present new, kpc-scale resolution data from the Atacama Large Millimeter/submillimeter Array (ALMA) and the Karl G. Jansky Very Large Array (VLA) that target line-emission from carbon monoxide (6-5), as well as continuum emission from dust and synchrotron radiation \citep[see also][]{lebowitz19}. We also present data from the Keck Cosmic Web Imager (KCWI) of the hydrogen recombination-line Ly$\alpha$, as well as other rest-frame ultra-violet emission lines that trace the ionized gas. This allows us to map in detail the radio source and host galaxy environment, in order to study the unusual properties of this high-$z$ radio galaxy. We will also investigate what drives the rapid displacement of molecular gas in this system: tidal forces from merging, or outflows driven by the radio jets and/or star formation. 
	
Throughout this paper, we assume the following cosmological parameters: $H_{0} = 71$\,\kmsMp, $\Omega_{\rm M} = 0.3$, and $\Omega_{\rm \Lambda} = 0.7$, which at $z = 1.9212$ (below) corresponds to an angular scale of 8.3 kpc arcsec$^{-1}$ and a luminosity distance $D_{\rm L} = 14.6\ {\rm Gpc}$ \citep{wright06}.

\section{Methods}

\subsection{ALMA}
\label{sec:alma}

The ALMA Cycle 4 observations (project 2016.1.01417.S) were conducted on 9 and 17 August 2017 for 1.2 hours on-source with 45 antennas with baselines up to $3.5\ {\rm km}$. We configured the spectral windows to cover two 4 GHz bands, one including the redshifted CO(6-5) line (235.8 -- 239.6\,GHz) and the other including only continuum (251.2 -- 255.0\,GHz). 

    The data were calibrated using the ALMA calibration pipeline that is included in the Common Astronomical Software Applications (CASA) version 4.7.2 \citep{casateam22}, by means of running the script that was supplied with the data by the North American ALMA Science Center (NAASC). Subsequent imaging was done manually using CASA version 5.3.0. For the line data, we subtracted the continuum in the $(u,v)$-domain by fitting a straight line to the line-free channels. We then combined our Cycle 4 data with 6 min of ALMA Cycle 2 observations taken at the same frequency but in a more compact configuration (baselines $\ge$17m; Paper {\sc II}). Imaging and deconvolution were done using a robust weighting scheme \citep{briggs95}, resulting in an image resolution of $0.12\arcsec \times 0.08\arcsec$ (1.0\,$\times$\,0.7 kpc) at a position angle (PA) 80$^{\circ}$. The line data were imaged with a spectral resolution of 15 km\,s$^{-1}$ for a single channel, resulting in a root-mean-square (rms) noise of 0.24 mJy\,beam$^{-1}$\,chan$^{-1}$. We also produced lower resolution line-data cubes by tapering the data to a beam-size of $0.23\arcsec \times 0.19\arcsec$ (1.9\,$\times$\,1.6 kpc) at PA $= 81^{\circ}$, with channels of 45 kms\ and an rms\ noise level of 0.15 mJy\,beam$^{-1}$\,chan$^{-1}$.

\subsection{VLA}
The VLA observations were conducted on 29 May 2015 in BnA-configuration, 5 and 13 August 2015 in A-configuration, and 30 December 2017 in B-configuration (projects VLA/15A-316 and VLA/17B-444). The total on-source time was 3.3 hours. The observations were centred at 43\,GHz and used an effective bandwidth of 7.5\,GHz. We used calibrator J2253+1608 for calibrating the bandpass response, J0204-1701 at 4.4$^{\circ}$ distance from our target for calibrating the complex gains every 60-70 seconds, and 3C147 for applying the absolute flux scale.

A standard manual data reduction and analysis was performed using CASA version 5.1.1 for calibrating the A- and BnA-configuration data, and CASA version 5.3.0 for calibrating the B-configuration data and subsequent imaging of the combined data. We used the multi-frequency synthesis method with a robust weighting scheme to create a continuum image with a resolution of 0.08$^{\prime\prime}$\,$\times$\,0.05$^{\prime\prime}$ (0.7\,$\times$\,0.4 kpc) at PA $= -6.6^{\circ}$. The data were cleaned, but the signal was not strong enough to perform a self-calibration. This means that low-level artifacts due to phase errors persist in the image. The rms noise level of the VLA continuum image is 24\,$\mu$Jy\,beam$^{-1}$.  

\subsection{Keck}
The Keck observations (project U074) were conducted on 6 October 2018 using the Keck Cosmic Web Imager at central wavelength 420\,nm \citep{morrissey18}. KCWI is an integral field unit (IFU) that effectively provides a spectrum at every spatial pixel (spaxel). The KCWI large slicer and BL grating was used to cover the effective wavelength range of 345 -- 525\,nm, or 118 -- 180\,nm in the rest frame, with a pixel size of $1.4\arcsec$ (12 kpc) in the E-W direction and $0.29\arcsec$ (2.4 kpc) in the N-S direction. The spectral resolution is R\,$\sim$\,900. The total exposure time was 1 hour, divided into six exposures of 10 minutes each. The data were reduced following standard procedures using the KCWI data reduction pipeline \citep{neill18}.

\section{Results}

\subsection{Molecular Gas and Dust}
\label{sec:gasdust}

\begin{figure*}
  \epsscale{1.0}
  \plotone{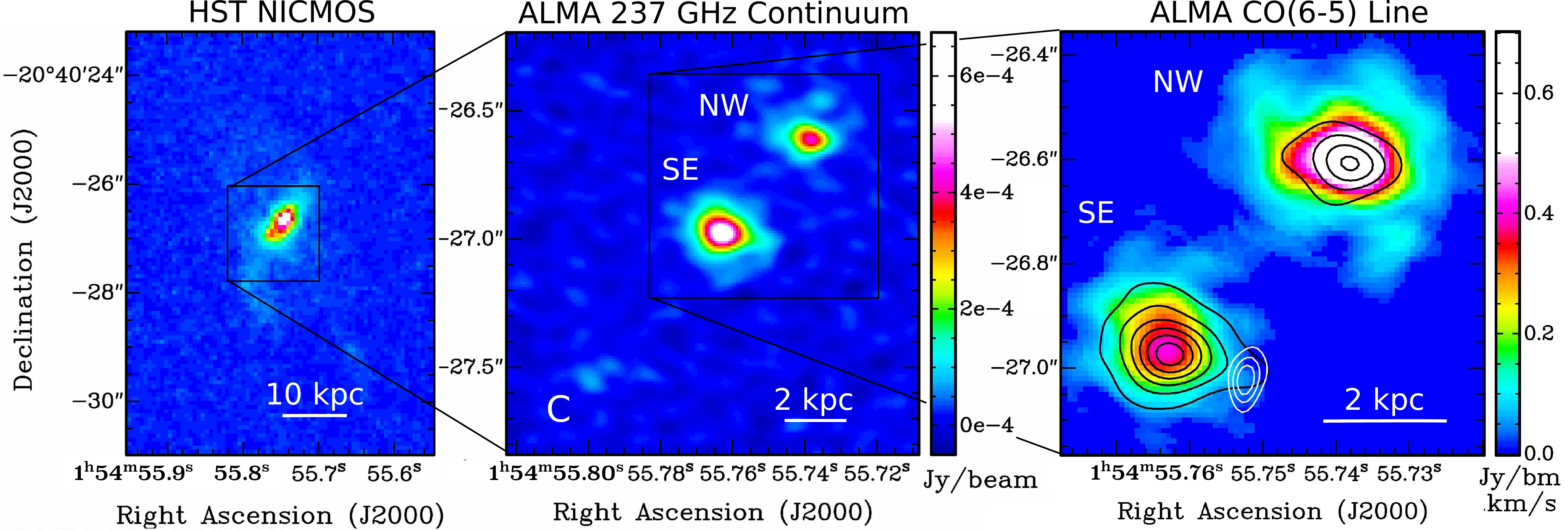}
\caption{Left: HST/NICMOS F160W Image of Dragonfly Galaxy $(z=1.92$) \citep{pentericci01}, Middle: 237\,GHz (692\,GHz rest frame) ALMA continuum emission in the inner 13 kpc. Right: Total Intensity map of ALMA CO(6-5) emission in the inner 7 kpc. The black contours are the 237\,GHz ALMA continuum from the middle plot. Contour levels start at 0.1 mJy\,beam$^{-1}$ and increase in steps of 0.1 mJy\,beam$^{-1}$. The white contours are the 43 GHz (115 GHz rest frame) VLA synchrotron emission of a radio hot spot (see Fig.~\ref{fig:vla} and Sect. \ref{sec:vla} for more details on the radio source). Contour levels start at 0.4 mJy\,beam$^{-1}$ and increase in steps of 0.4 mJy\,beam$^{-1}$.}
\label{fig:dragonfly}
\end{figure*}

\begin{figure*}
  \epsscale{1.0}
\plotone{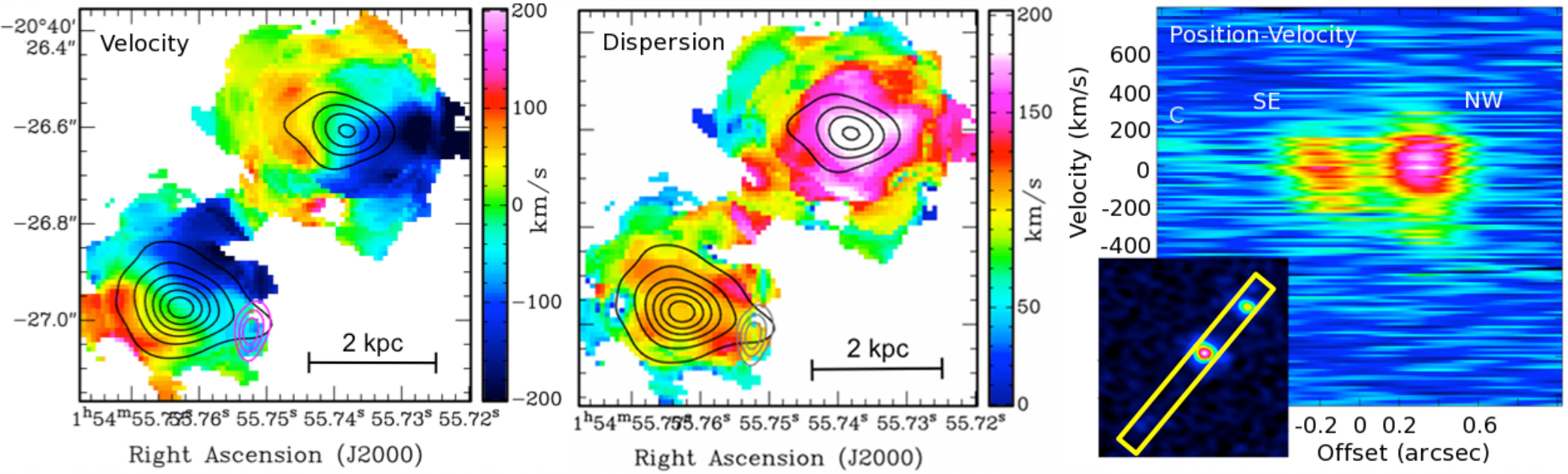}
\caption{Left: Moment 1 Velocity Map of CO(6-5), Middle: Moment 2 Velocity Dispersion Map of CO(6-5). The contours show the ALMA and VLA continuum data from Fig.~\ref{fig:dragonfly}. Right: Position Velocity Map (smoothed to resolution of 0.3$^{\prime\prime}$) taken along the position angle shown as yellow rectangle in the inset on the bottom left.}
\label{fig:momentmaps}
\end{figure*}

Figure~\ref{fig:dragonfly} (left) shows an HST image of the Dragonfly Galaxy \citep{pentericci01}. On a scale of several tens of kpc, the HST image shows prominent tidal features that were previously described in Paper {\sc I}. Figure~\ref{fig:dragonfly} (middle) shows the ALMA 237\,GHz continuum image, which reveals three components. Components NW and SE appear in the center of the Dragonfly system and were previous detected in Paper {\sc II}. The continuum flux in component SE is roughly twice that of component NE (Paper {\sc II}), confirming that the SE galaxy hosts the bulk of the dust-obscured star formation detected by \citet{drouart14} and \citet{falkendal19}. Component C is a likely companion galaxy that was previously detected in CO(6-5) but not the 237\,GHz continuum (Paper {\sc II}). These new observations reveal a peak flux density of 70\,$\mu$Jy\,beam$^{-1}$ for the observed 237 GHz continuum of component C. 

Figure~\ref{fig:dragonfly} (right) shows a total intensity map of the CO (6-5) emission in the central $\sim$8 kpc. The ALMA data clearly show that the Dragonfly system consists of two central galaxies, NW and SE, which are separated by approximately 4~kpc. The high-$J$ line of CO(6-5) traces molecular disks that have a projected radius of approximately 2~kpc for NW and 1.5 kpc for SE. Component C could be a third galaxy (see also Paper {\sc II}).

Figure~\ref{fig:momentmaps} (left) shows a moment-1 velocity map of CO(6-5). Both NW and SE possess distinct regions of redshifted and blueshifted gas, indicative of rotating disks. The disk of the NW galaxy shows less ordered gas kinematics and a higher velocity dispersion compared to the disk of the SE galaxy. The two disks are connected by a tidal bridge of gas. This bridge of CO emission between the galaxies is also visible in the position-velocity plot with the pseudo-slit taken along the radio axis (Fig.~\ref{fig:momentmaps} - right). 

Figure \ref{fig:momentmaps} (middle) shows the moment-2 map of the CO velocity dispersion, which is defined as $\sigma_{\rm CO}$\,=\,$FWHM_{\rm CO}$/2.355, with $FWHM_{\rm CO}$ the full width at half the maximum intensity of the CO line. Figure \ref{fig:momentmaps} (middle) shows that the molecular gas in the bridge has a velocity dispersion that ranges from $\sigma_{\rm CO}$\,=\,90\,$-$\,170\,km\,s$^{-1}$, where the maximum $\sigma_{\rm CO}$\,=\,170\,km\,s$^{-1}$ is higher than the velocity dispersion seen throughout the disk of the SE galaxy. Based on the 8.2 GHz radio continuum image from \citet{pentericci00}, in Paper {\sc II} it was argued that that the NW galaxy must host the radio-loud AGN.

\subsection{Radio-loud AGN}
\label{sec:vla}

\begin{figure}
\epsscale{1.15}
\plotone{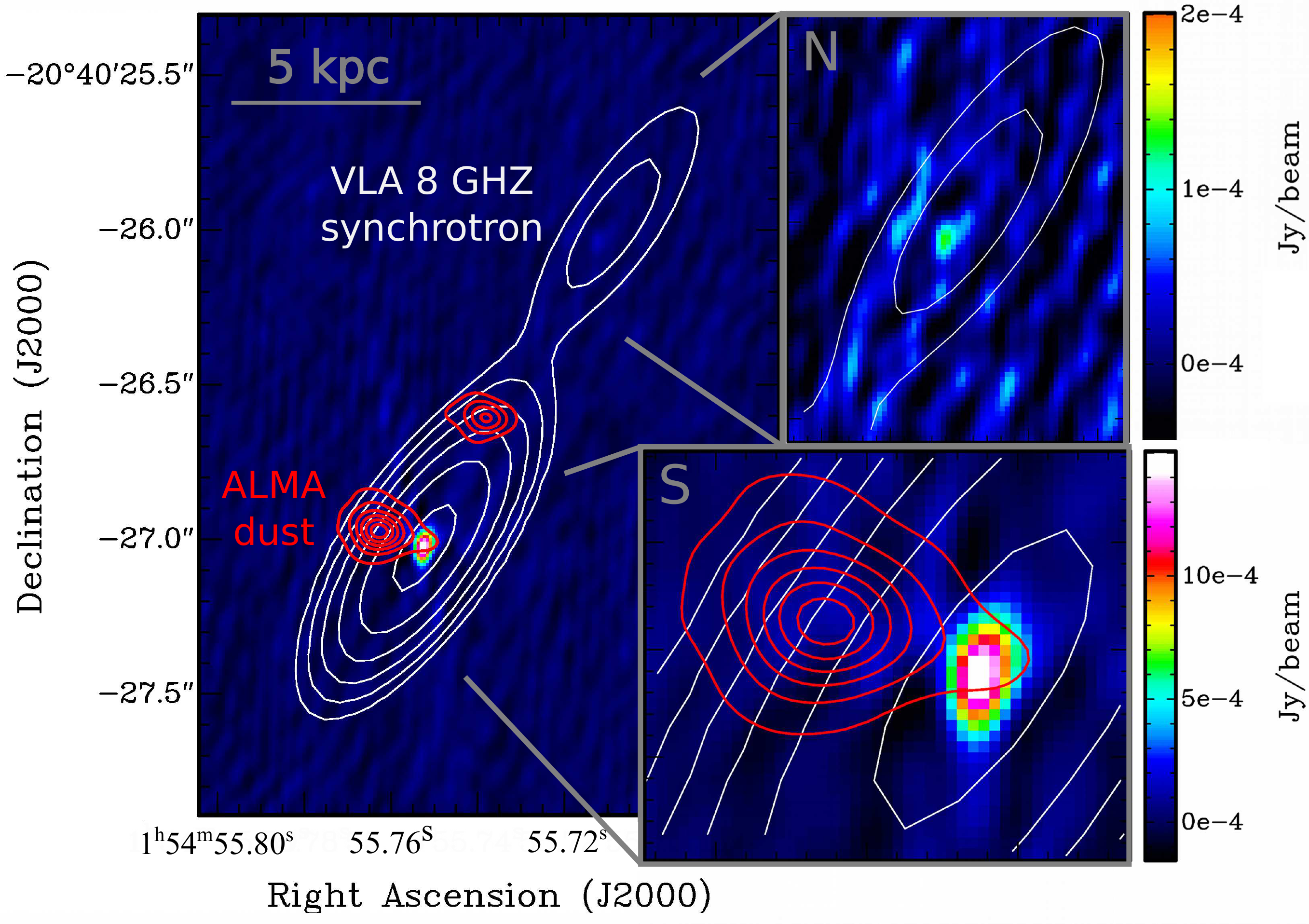}
\caption{VLA 43 GHz continuum image of the radio source, with overlaid previous lower resolution VLA data of the radio source at 8 GHz \citep[white contours;][]{pentericci00}. The VLA 8 GHz contours start at 1 mJy\,beam$^{-1}$ and increase by a factor of 2.  Self-calibration of the 8 GHz data could have affected its positional accuracy, so we shifted the 8 GHz radio continuum by 0.09 arcsec to create the best possible overlay with the 43 GHz continuum. The red contours show the ALMA 237\,GHz dust continuum from Fig.~\ref{fig:dragonfly}. The inset on the bottom-right shows the bright 43\,GHz hot-spot in the southern lobe, detected on the right edge of the disk of the SE galaxy. The ALMA dust continuum may be contaminated by faint 237\,GHz synchrotron emission at the location of the hot-spot. The inset on the top-right shows a weak and unresolved 43\,GHz counterpart to the northern lobe.}
\label{fig:vla}
\end{figure}

The radio source that classifies the Dragonfly Galaxy as a high-redshift radio galaxy is shown in Figure~\ref{fig:vla}. The contours of the radio jet detected by the VLA at 8\,GHz \citep{pentericci00} are overlaid on our new high-resolution VLA image at 43\,GHz. The alignment of the central axis of the 8\,GHz radio source with the NW galaxy suggests that NW is the host galaxy of the radio source (Paper {\sc II}). The two-sided morphology of the radio source is consistent with our new VLA observations. At 43 GHz, the high resolving power of the VLA left too little surface brightness sensitivity to pick up any radio lobes associated with a jet, but a bright spot was detected on the western edge of the SE galaxy. This bright spot is aligned along the radio axis and has a flux density of $S_{43}$\,=\,1.8\,$\pm$0.2 mJy. The northern 8\,GHz lobe shows a weak and apparently unresolved 43\,GHz counterpart at a 5.5$\sigma$ level, with a peak flux density of $S_{43}$\,=\,0.13\,$\pm$0.5 mJy\,beam$^{-1}$. The projected distance between the northern and southern radio component detected at 43\,GHz is $1.15\arcsec$, or 9.5\,kpc. A central radio core is not detected in our VLA data.

Previous low-resolution observations performed with the Australia Telescope Compact Array at 39.9 GHz revealed a continuum flux density of $S_{39.9}$\,=\,{\bf 5.1}\,$\pm$\,1.5 mJy\,beam$^{-1}$ with a spectral index of $\alpha$\,=\,-1.3, for $S_{\nu}$\,$\propto$\,$\nu^{\alpha}$ \citep{emonts11}. These observations had a spatial resolution of 9.8$^{\prime\prime}$\,$\times$7.1$^{\prime\prime}$ (81\,$\times$\,59 kpc), hence did not resolve any of the structures associated with the Dragonfly system. Based on these previous observations, we expect a total 43 GHz continuum flux density of  $S_{43}$\,$\sim$\,4.6 mJy\,beam$^{-1}$. This means that we only recover about fourty percent of the 43\,GHz continuum emission in our high-resolution VLA data. The remaining sixty percent is likely synchrotron emission from more diffuse lobes that are resolved out in our high-resolution VLA data.

\subsection{Keck Spectroscopy}
\begin{figure}
\epsscale{1.15}
\plotone{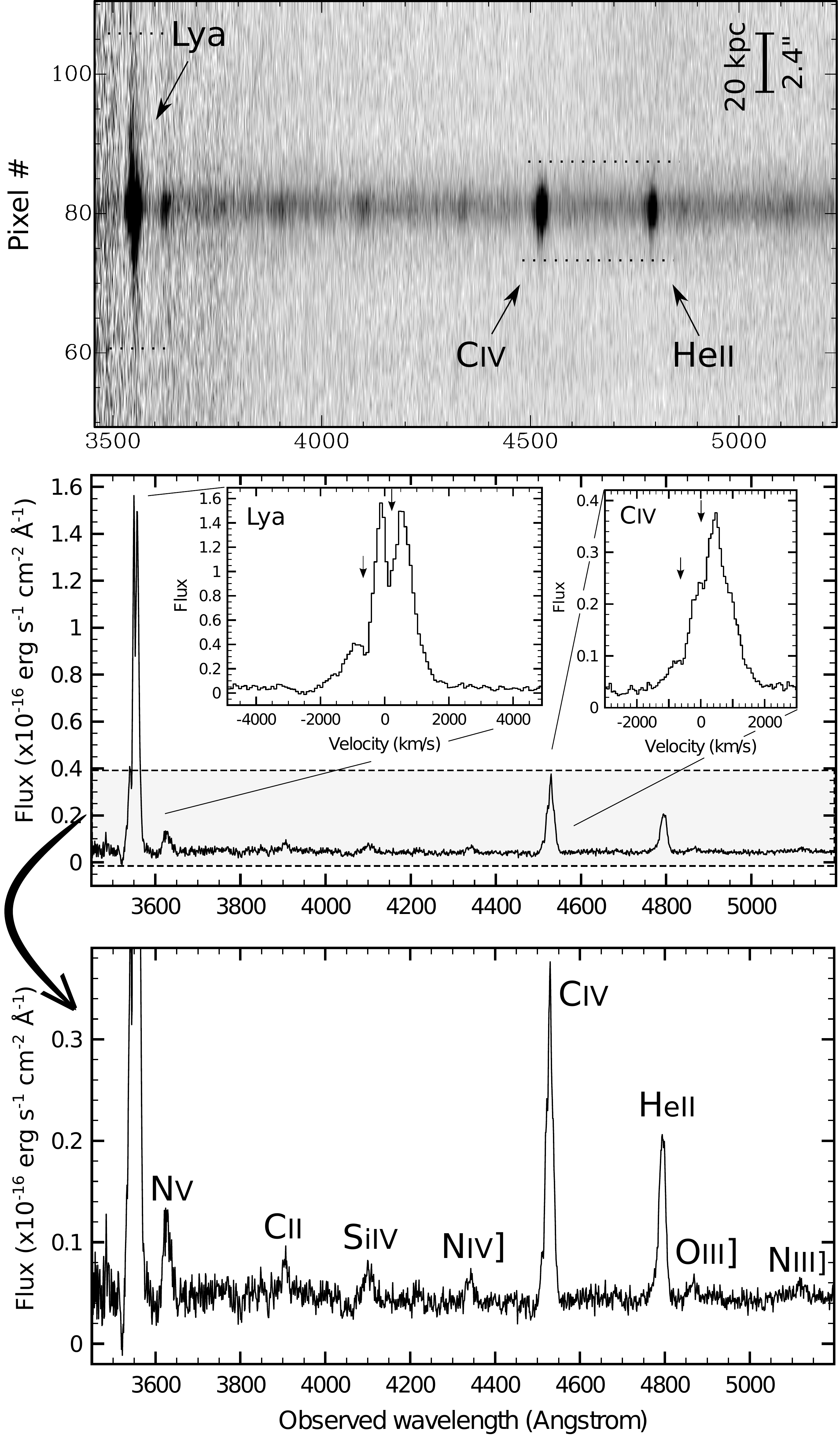}
\caption{Keck spectroscopy. Top: 2D spectrum taken with a $6.8\arcsec$ wide pseudo-slit along the North-South direction and covering both galaxies. The pixel-scale on the vertical axis is $0.29\arcsec$ per pixel. The dotted lines mark the approximate spatial extent of the Ly$\alpha$, \CIV\ and \HeII\ lines. Middle+bottom: 1D spectrum extracted against $9 \times 2$ spaxels in the region of the Dragonfly. Visible are the lines: Ly$\alpha$ ($\lambda_{\rm rest} = 1215.7$\,\AA), \NV\ ($\lambda_{\rm rest} = 1240$\,\AA), \CII\ ($\lambda_{\rm rest} = 1336$\,\AA), \SiIV\ ($\lambda_{\rm rest} = 1400$\,\AA), \NIV] ($\lambda_{\rm rest} = 1486$\,\AA), \CIV\ ($\lambda_{\rm rest} = 1548+1551$\,\AA), \HeII\ ($\lambda_{\rm rest} = 1640$\,\AA), \OIII] ($\lambda_{\rm rest} = 1663$\,\AA), and \NIII] ($\lambda_{\rm rest} = 1750$\,\AA). The middle panel also shows a zoom-in of the Ly$\alpha$ (left) and \CIV\ (right) lines, with arrows indicating potential absorption features. The zero velocity for the close doublet of \CIV\ is centered on $\lambda_{\rm rest} = 1548.2$\,\AA.} 
\label{fig:keck}
\end{figure}

\begin{figure*}
\epsscale{1.1}
\plotone{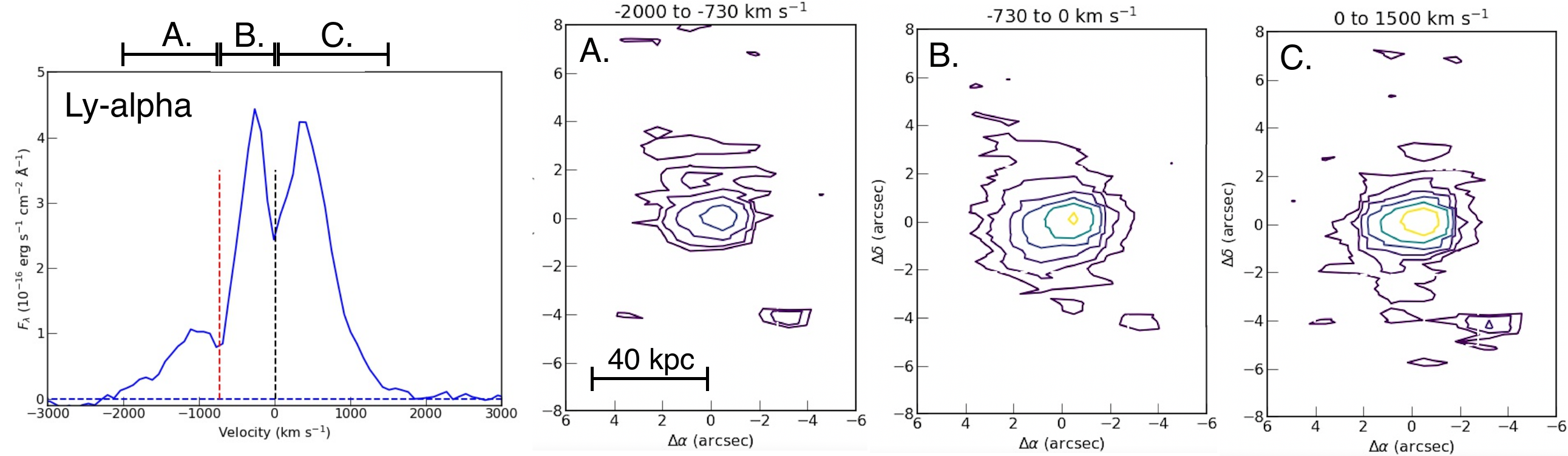}
\caption{Left: Ly$\alpha$ spectrum of the Dragonfly Galaxy. A velocity correction was applied to account for the Earth's rotational and orbital motion, and the motion of the Galaxy with respect to the Cosmic Microwave Background (CMB) frame. The 0-velocity (black dashed line) corresponds to $z$\,=\,1.9212. The red dashed line highlights $v$\,=\,-750 km\,s$^{-1}$. Annotated on the top are three velocity ranges, from -2000 to -730 km\,s$^{-1}$ (A), -730 to 0 km\,s$^{-1}$ (B), and 0 - 1500 km\,s$^{-1}$ (C). Right: Three contour plots of the average surface brightness of the Ly$\alpha$ flux across the velocity ranges A, B, and C. In each plot, contours start at 10$^{-17} \, \rm erg\,s^{-1}\,cm^{-2}$\AA$^{-1}$ 
and increase by a factor of two. The center at (0,0) is the intensity weighted centroid of the Ly-alpha line in the Keck data. The PSF of the core of the Ly$\alpha$ line has $FWHM$\,=\,1.3 arcsec in the North-South dimension and $FWHM$\,=\,2.2 arcsec in the East-West dimension.}
\label{fig:keckmaps}
\end{figure*}

Figure \ref{fig:keck} shows a 2D spectrum of the Keck data, obtained by putting a `pseudo-slit' with a width of 5 spaxels (6.75 arcsec) aligned in north-south direction across the Dragonfly Galaxy. We produced this `pseudo-slit' from the IFU data through a linear arrangement of spaxels. The large spaxel size of the KCWI data means that the NE and SW galaxies cannot be accurately distinguished in the Keck imaging. We detect in total nine emission lines, including Ly$\alpha$, \ion{N}{5}, \ion{C}{2}, \ion{Si}{4}, \ion{N}{4}], \ion{C}{4}, \ion{He}{2}, \ion{O}{3}], and \ion{N}{3}]. Based on the Keck spectrum, we can classify this AGN as Type II. Type II AGN are characterized by a core that is blocked by the torus, producing only narrow emission lines and lacking the very broad lines from the broad-line region that are typically seen in Type I AGN. 

Figure \ref{fig:keck} shows that the brightest emission lines (Ly$\alpha$, \ion{He}{2}, and \ion{C}{4}) appear spatially extended. The Ly$\alpha$ emission stretches across a scale of at least 100 kpc. This is analogue to spectroscopic results of giant Ly$\alpha$ halos observed in other high-$z$ radio galaxies \citep[e.g.,][]{ojik96, pentericci97, villar03, villar07, vernet17} and quasars \citep[e.g.,][]{cantalupo14,borisova16}. The Ly$\alpha$ line also shows two putative absorption features, one around the redshift of CO, and the other blueshifted by roughly 600 km\,s$^{-1}$. These absorption systems are tentatively detected also in \ion{C}{4}. Similar absorption features were seen in other high-$z$ radio galaxies \citep[e.g.,][]{jarvis03,humphrey08, swinbank15, kolwa19}. Emission of the \ion{He}{2} and \ion{C}{4} metal lines is visible across approximately 35 kpc (Fig.\,\ref{fig:keck}, top panel). 

Figure \ref{fig:keckmaps} shows that the widespread Ly$\alpha$ emission is bright enough that it can be imaged across $\sim$80 kpc. At the lowest velocities, the extended Ly$\alpha$ emission is found predominantly North to North-East of the galaxies, while at the highest velocities it is found stretching towards the South and South-West. Ly$\alpha$ emission at larger scales or extended emission seen in the other lines (as shown in the 2D spectrum of Fig. \ref{fig:keck}) has too low surface brightness to be visible above the noise in similar contour plots.

Previously, also CO(1-0) emission from cold molecular gas was observed across $\sim 60$ kpc, predominantly at positive velocities (Paper {\sc I}). Our new Keck results add to the evidence that there is a rich circumgalactic medium present around the Dragonfly system, although a full analysis of the circumgalactic gaseous environment of the Dragonfly is beyond the scope of the current paper.

\section{Discussion}

\subsection{Merging disk galaxies}
\label{sec:disks}

The ALMA CO(6-5) moment maps show that both the SE and NW components are likely rotating disk galaxies. The galaxies are connected by a tidal bridge of molecular gas. The presence of this tidal gas connecting NW and SE, along with their small separation ($\sim 4$ kpc), indicates that these galaxies are in an early stage of merging \citep[e.g.,][]{mihos96,prochaska19}. In agreement with this, studies at low redshift showed that radio-AGN activity can be triggered during the pre-coalescent phase of a merger event \citep{ramos11,tadhunter11}. The center of the bridge of molecular gas between the two galaxies shows a high velocity dispersion, as well as a sudden change in gas kinematics. This could indicate that in the middle of the bridge the gas is strongly shocked, which could convert the bulk velocity into a turbulence \citep[e.g.,][]{cornuault18}. Our results are in good agreement with simulations by \citet{sparre21}, which reveal that a major merger between two disk galaxies can create a gas bridge that is dominated by supersonic turbulence at the midpoint in between the galaxies. This bridge can form either before or after the first passage of the galaxies.

The center of the NW galaxy contains the highest velocity dispersion in this system and has more chaotic rotation than the SE galaxy. The maximum velocity dispersion of $\sigma_{CO}$\,$\sim$\,200 km\,s$^{-1}$ in the central region of the NW galaxy is significantly higher than the velocity dispersion typically observed in both low- and high-$z$ systems \citep[e.g.,][]{swinbank11,girard21}. Following \textcolor{blue}{Sect. \ref{sec:vla}}, the NW galaxy hosts the radio-loud AGN. We speculate that the AGN activity may be related to the high velocity dispersion of the gas in the NW galaxy, either because the radio-loud AGN influences the gas that is present in the NW galaxy \citep[e.g.,][]{morganti21,ramos-almeida22}, or because large amounts of molecular gas are being accreted onto the NW galaxy, which may fuel the AGN activity \citep[e.g.,][]{ruffa19}. Alternatively, we cannot rule out that the disturbed rotational kinematics of the NW galaxy are caused by another, much closer galaxy-galaxy interaction associated with only the NW component, rather than a rotating disk. \citet{hibbard96} show examples of nearby galaxy mergers where gas kinematics leave blue- and redshifted signatures that could be difficult to distinguish from rotating disks if these systems were to be located at larger distances. However, we argue that this is unlikely, because the HST and ALMA continuum imaging do not reveal indications that the NW galaxy consists of multiple component.

\subsection{Gas displacement: tidal debris or outflow?} 
 
Previous ALMA observations of the CO(6-5) emission at lower spatial resolution revealed that molecular gas is rapidly being displaced within the Dragonfly galaxy, with gas displacement rates between 1200 -- 3000 ${\rm M}_{\odot}\,{\rm yr}^{-1}$ (Paper {\sc II}). Because this is similar to the star formation rate of $SFR$\,$\sim$\,2000 M$_{\odot}$\,yr$^{-1}$, the mechanism that displaces the molecular gas is likely critical for the evolution of this system. Paper {\sc II} could not clearly distinguish whether this mechanism was related to gravitational interaction between two disks, AGN- or starburst-driven outflows, or both. The bridge of CO(6-5) that we detect in our new ALMA data is a clear indication that gravitational effects of the merger are important, but the bridge likely represents only a fraction of the total gas displacement in this system. To further investigate the gas displacement in the Dragonfly Galaxy, we tapered our ALMA CO(6-5) data to a lower resolution of $0.23\arcsec \times 0.19\arcsec$ (1.9\,$\times$\,1.6 kpc; see Sect. \ref{sec:alma}) in order to increase our sensitivity for tracing molecular gas at lower surface brightness levels. 
 
\subsubsection{Tidal debris}
\label{sec:tidal}
 
\begin{figure*}
\plotone{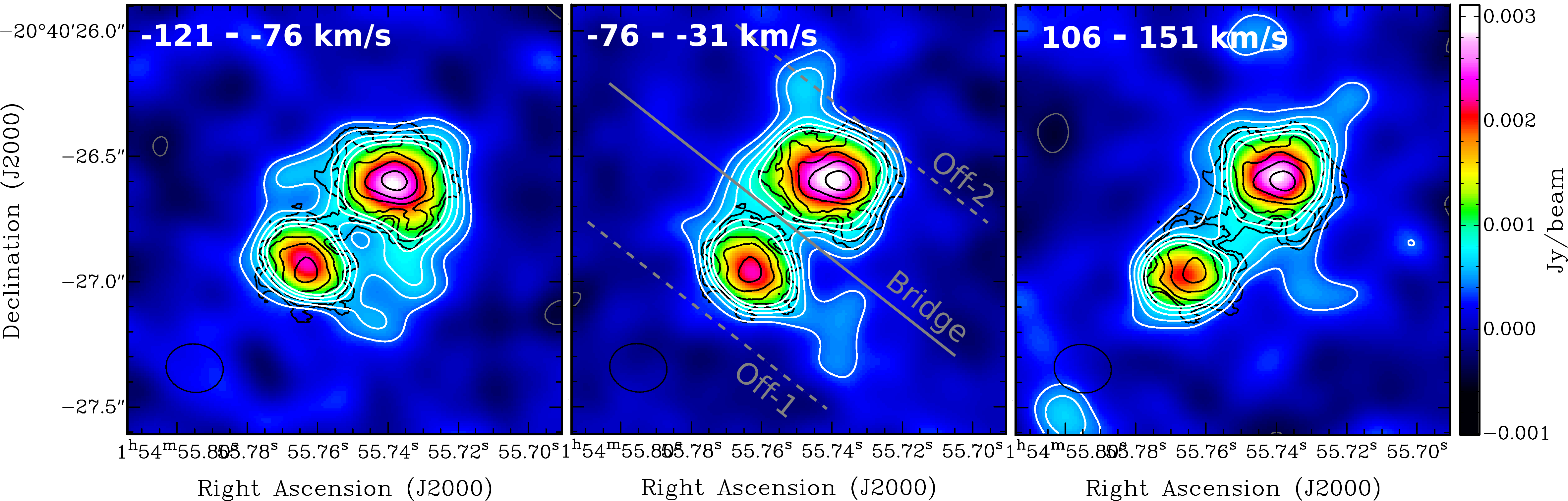}
\caption{CO(6-5) emission at various velocities after tapering our ALMA data to a lower resolution of 0.23\,$\times$\,0.19$^{\prime\prime}$ (1.9\,$\times$\,1.6~kpc) at PA 81$^{^\circ}$. White contours mark the level of the integrated emission after binning the tapered data to 45 km\,s$^{-1}$ channels. Contour levels start at 2.5$\sigma$ and increase by 1$\sigma$, with $\sigma$\,=\,0.15 mJy\,beam$^{-1}$ (corresponding negative contours are in grey). The black contours show the CO total intensity image at full spatial resolution from Fig.~\ref{fig:momentmaps}. Contour levels start at 0.02 Jy\,beam$^{-1}$\,$\times$\,km\,s$^{-1}$ and increase by factor 2. The three lines in the middle panel visualize the axes along which we obtained the position-velocity (PV) plots to estimate the CO luminosity of the bridge material. These PV plots are shown in Fig.~\ref{fig:bridge}.}
\label{fig:chanmaps}
\end{figure*}

In addition to the relatively narrow CO bridge between the two galaxies, our tapered ALMA data show additional CO(6-5) emission at lower surface brightness levels at velocities $|v| \la 250\,{\rm km}\,{\rm s}^{-1}$ (Fig.~\ref{fig:chanmaps}). This emission is also seen in the position-velocity plot of Fig.~\ref{fig:momentmaps}. Spectra of the CO(6-5) data reveal no kinematic evidence for outflows in the region between the galaxies, so we conclude that the molecular gas in between the galaxies is mostly tidal debris from the gravitational interaction between the gaseous disks of the two galaxies. Simulations by \citet{sparre21} show that, in addition to the formation of a gaseous bridge, a major merger may lead to a substantial mass accretion from entrainment and cooling of the circumgalactic medium. This process may contribute to the tidal material the we find between the galaxies in the Dragonfly system.

\begin{figure*}
\plotone{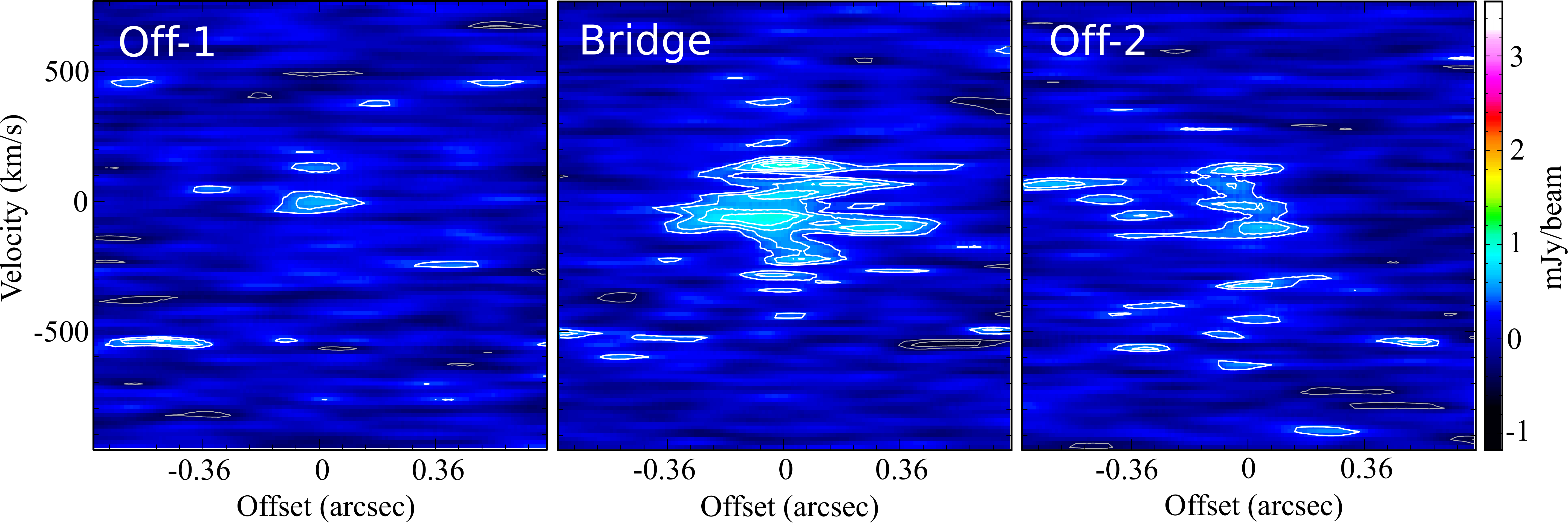}
\caption{Position-Velocity plots of the CO(6-5) emission in the tapered data taken across the bridge between the two galaxies (middle), and two `off' positions (left and right), as indicated by the lines in the middle panel of Fig.~\ref{fig:chanmaps}. The `off' slits are at the same (but opposite) distance from the center of the SE (Off-1) and NW (Off-2) galaxy as the slit across the bridge. These `off' slits therefore  give an indication how much of the `bridge' emission is contaminated by emission from the bright central region of the galaxies due to the point-spread-function (i.e., synthesized beam) of our observations. Contour levels start at 2.5$\sigma$ and increase with $1 \sigma$, with $\sigma = 0.15\,{\rm mJy}\,{\rm beam}^{-1}$ (negative contours are in grey).} 
\label{fig:bridge}
\end{figure*}

To estimate the CO(6-5) luminosity of this tidal debris, we extract a pseudo-slit in perpendicular direction across the bridge, as shown in Fig.~\ref{fig:bridge}. The CO emission detected across the tidal debris is contaminated by emission from the bright CO in the NW and SE galaxy, as a result of the point-spread-function (PSF) of the synthesized beam. To estimate this PSF contamination, we assume that the molecular gas in the central disks of the two galaxies is distributed roughly symmetrically around the center of the galaxies. As shown in Fig.~\ref{fig:bridge}, we then take two pseudo-slits at `off' positions on opposite side of the NW and SE galaxy, at equal distance from the center of the galaxies as the pseudo-slit of the `bridge'. To estimate the true CO(6-5) luminosity of the tidal debris, we subtract the CO signal in regions `off-1' and `off-2' from the CO signal of the `bridge' region. The resulting CO(6-5) intensity of the molecular gas in the tidal debris, corrected for PSF contamination of the galaxies, is $I_{\rm CO(6-5)}$ = 0.3\,$\pm$\,0.1 Jy\,beam$^{-1}$\,$\times$\,km\,s$^{-1}$.
 
\subsubsection{AGN-driven outflow} 
 \label{sec:outflow}
 
\begin{figure*}
\epsscale{0.8}
\plotone{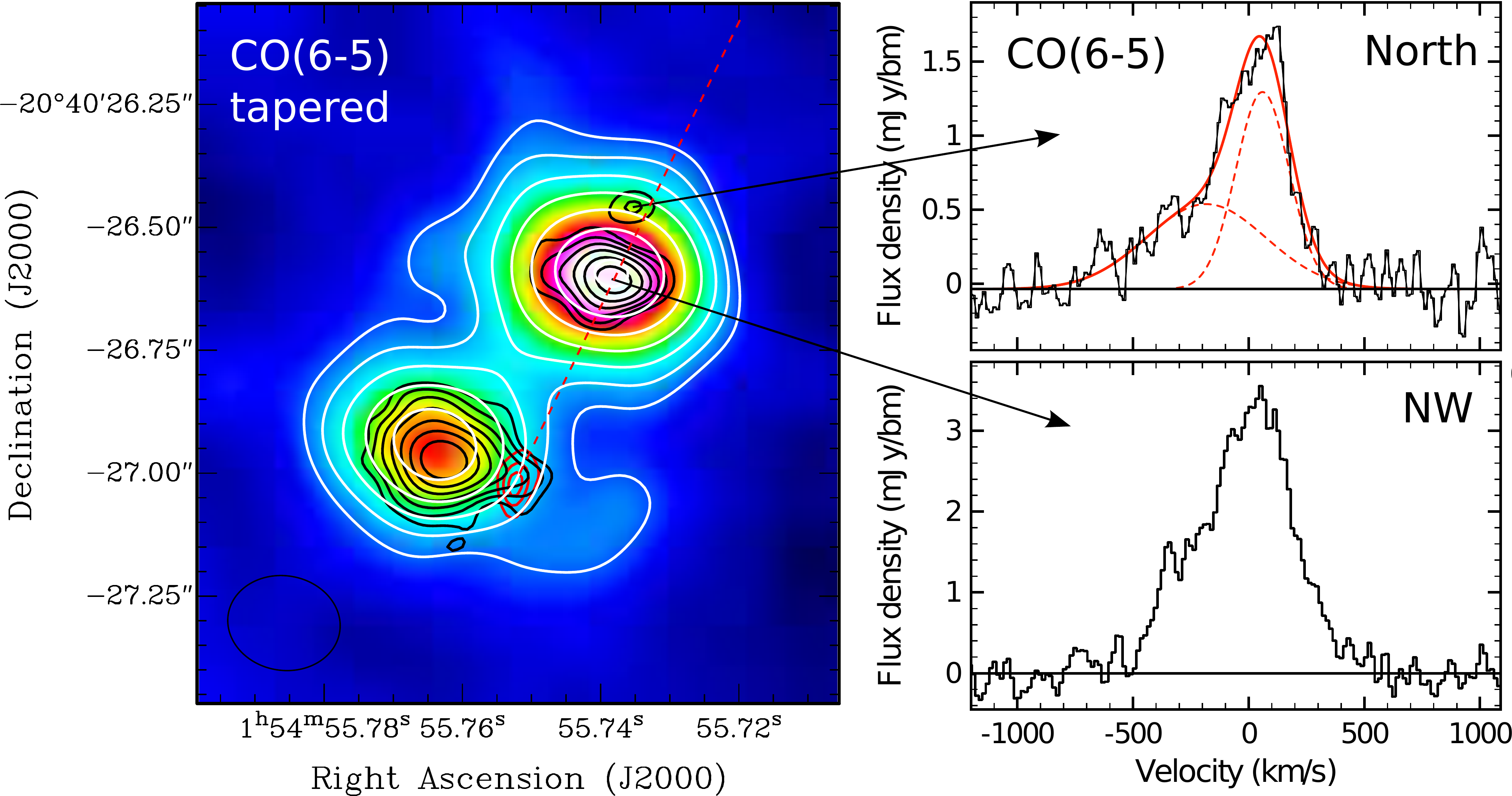}
\caption{CO(6-5) ALMA image and spectrum with data tapered to a lower resolution, as in Fig.~\ref{fig:chanmaps}. Left: Total intensity image of the CO(6-5) integrated across the velocity range from -750 to +400 km\,s$^{-1}$. White contours show the total intensity map starting at 4$\sigma$ and increasing by a factor $\sqrt{2}$, with $\sigma$\,=\,0.05 Jy\,bm$^{-1}$\,$\times$\,km\,s$^{-1}$. Black contours are the ALMA 237\,GHz continuum at full (untapered) resolution from Fig.~\ref{fig:dragonfly}, starting at 5$\sigma$ and increasing by a factor $\sqrt{2}$, with $\sigma$\,=\,15 $\mu$Jy\,bm$^{-1}$. Red contours are the VLA 43\,GHz continuum from Fig.~\ref{fig:vla} at levels 0.3, 0.8, 1.3 mJy\,bm$^{-1}$. The red dashed line visualizes the axis of the radio source. The synthesized beam of the tapered data is 0.23$^{\prime\prime}$\,$\times$\,0.19$^{\prime\prime}$ (1.9\,$\times$\,1.6~kpc) at PA 81$^{\circ}$, as indicated with the ellipse in the bottom-left corner of the image. Middle: spectra of the NW galaxy (bottom) and a region north of the NW galaxy that corresponds to a faint blob of ALMA continuum emission (top). For clarity, a Hanning smoothing has been applied to the spectra. The spectrum of the NW galaxy nucleus is complex (see Paper {\sc II}), while the spectrum north of the galaxy can be fitted with two Gaussian functions (red lines). The two spectra are not entirely mutually independent, but suggest the presence of an outflow component that peaks North of the nucleus along the jet-axis.}
\label{fig:COoutflow}
\end{figure*}

The CO emission at the highest velocities (|$v$|$\ga$250 km s$^{-1}$) appears to be localized around the NW galaxy. Figure \ref{fig:COoutflow} (middle) shows the CO(6-5) spectra from the tapered data against the central region of NW, as well as  a region 0.15$^{\prime\prime}$ (1.2~kpc) to the North along the axis of the radio source. These spectra appear to have a broad, blueshifted wing to the CO profile. This blue wing is indicative of an outflow from the NW galaxy. The maximum velocity of the outflow reaches from $-500 \rm \,km\,s^{-1}$ in the center to approximately $-800\,{\rm km}\,{\rm s}^{-1}$ North of the center. No outflow is seen against the SE galaxy, which hosts the bulk of the star formation.\footnote{\citet{spilker20a,spilker20b} argue from OH 119$\mu$m absorption measurements that molecular outflows driven by star formation are ubiquitous among high-$z$ dusty star-forming galaxies, but that molecular emission lines may not always be reliable tracers for these outflows.} This, together with the apparent orientation of the outflow along the jet axis, suggest that the outflow from the NW galaxy is most likely driven by the radio source. 

In the region North of the center of NW, where the outflow reaches its maximum velocity, we also detect a faint blob in the 237 GHz radio continuum, with a peak flux-density of roughly  0.09\,mJy\,beam$^{-1}$ (Fig.~\ref{fig:COoutflow}). Since we do not detect anything in the 43 GHz continuum at this location, we can set a 3$\sigma$ lower limit to the spectral index of $\alpha$\,>\,0.12, for  $S$\,$\propto$\,$\nu^{\alpha}$. This inverted spectral index makes it unlikely that this faint ALMA continuum detection is radio synchrotron emission from a hot-spot in the radio jet, but if real, its exact nature is not clear.

The Gaussian fitting to the CO profile in the region North of the center (Fig.~\ref{fig:COoutflow}) reveals that the maximum velocity offset between the peak of the broad and narrow CO component is $\Delta v$\,=\,240\,$\pm$\,60 km\,s$^{-1}$, with $FWHM_{\rm CO}$\,=\,610\,$\pm$\,90 km\,s$^{-1}$ for the broad component (see Table \ref{tab:outflow}). The total CO intensity of the outflow is $I_{\rm CO} = 0.37 \pm 0.10\,{\rm Jy}\,{\rm beam}^{-1} \times\,{\rm km}\,{\rm s}^{-1}$.

\begin{figure}
\epsscale{1.0}
\plotone{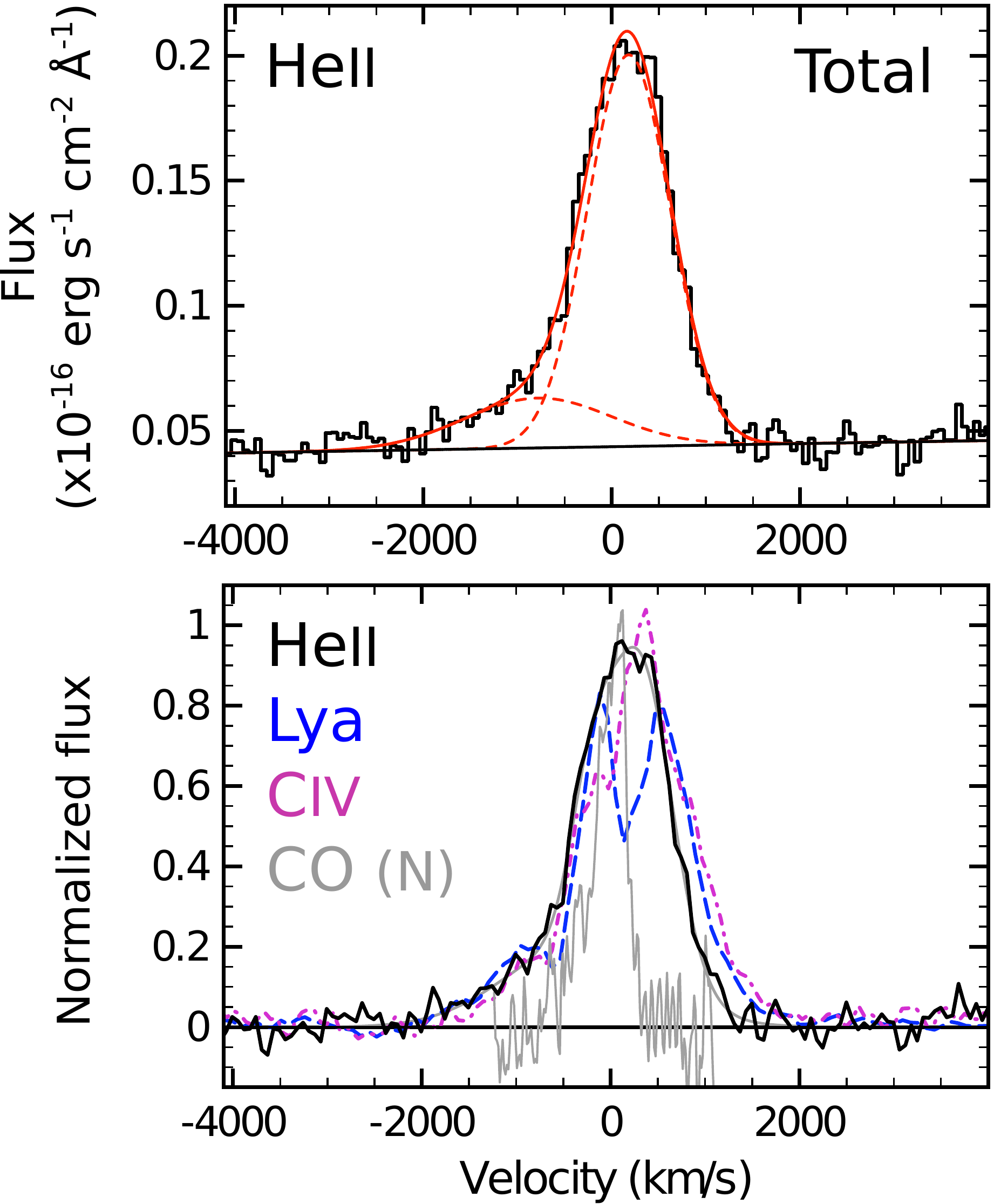}
\caption{Gas outflow in the optical emission lines. Top: Keck spectrum of \HeII\, fitted with two Gaussian components (red dashed lines). The spectrum is from Fig.~\ref{fig:keck} and does not spatially resolve the SE and NW galaxies. Bottom: Superposition of the profiles from \HeII\ (black solid), Ly$\alpha$ (blue dashed), \CIV\ (magenta dash-dotted), and CO(6-5) (grey thin). The optical profiles show the total line flux integrated across NW and SE (Fig.\,\ref{fig:keck}), while the CO profile shows only the emission against the Northern region from Fig.\,\ref{fig:COoutflow}. Fluxes have been normalized for easy comparison.}
\label{fig:opticaloutflow}
\end{figure}

The optical emission lines in the Keck spectrum also reveal a broad, blueshifted component (Fig.~\ref{fig:opticaloutflow}). The He\,{\sc II} line is a non-resonant line, and reveals a broad component that has a $FWHM$\,$\sim$\,1810\,$\pm$\,320 km\,s$^{-1}$, with a peak that is blueshifted by $\Delta v\,=\,-960\,\pm$\,220 km\,s$^{-1}$ with respect to the narrow He\,II component (Table \ref{tab:outflow}). As shown in Fig.\,\ref{fig:opticaloutflow}, Ly$\alpha$ and \ion{C}{4} show similar kinematics. Although the spatial resolution of these Keck data is not sufficient to determine the location of this broad component, it nevertheless suggests that the outflow seen in CO has a component also in the warm ionised gas. Moreover, the outflow velocity is more extreme in the ionised gas (Fig.~\ref{fig:opticaloutflow}).

\subsubsection{Mass displacement rates}

To derive molecular gas masses for the material in the bridge and the outflow, we first convert the derived CO(6-5) intensities from Sect.\,\ref{sec:tidal} and \ref{sec:outflow} to the expected intensity of the CO ground transition. For this, we assume the same CO(6-5)/CO(1-0) line ratios observed in Paper {\sc II}, namely $R_{I(6\rightarrow1)}$\,=\,$I_{\rm CO(6-5)}$/$I_{\rm CO(1-0)}$\,$\sim$\,13 for the molecular gas at |$v$|$\la$250 km s$^{-1}$ (i.e., the tidal debris), and 17\,$\la$\,$r_{\rm 6-5}$\,$\la$\,36 for the molecular gas at |$v$|$\ga$250 km s$^{-1}$ (i.e., the outflow). This allows us to estimate the luminosity of the CO ground-state transition, $L_{\rm CO(1-0)}^{\prime}$, following \citet[][and references therein]{solomon05}:
\ \\
\begin{equation}
L'_{\rm CO{\bf (1-0)}} = 3.25 \times 10^7 \left(\frac{I_{\rm CO(6-5)}}{R_{I(6\rightarrow1)}}\right) D_{\rm L}^2 ~ \nu_{\rm rest}^{-2} \left(1+z\right)^{-1},
\label{eq:lco}
\end{equation}
\ \\
with $I_{\rm CO(6-5)}$ the CO(6-5) intensity in Jy\,beam$^{-1}$\,$\times$\,km\,s$^{-1}$, $D_{\rm L}$ the luminosity distance in Mpc, $\nu_{\rm rest}$ the rest frequency of the CO(1-0) line in GHz (115.27\,GHz), and $L_{\rm CO(1-0)}^{\prime}$ the CO(1-0) luminosity in K~km\,s$^{-1}$~pc$^{2}$.
To derive molecular gas masses, we then use the same conservative CO-to-H$_{2}$ conversion factor of $\alpha_{\rm CO}$\,=\,$M_{\rm H2}$/$L^{\prime}_{\rm CO(1-0)}$\,=\,0.8 M$_{\odot}$ (K km\,s$^{-1}$ pc$^{2}$)$^{-1}$ as was used in Paper II, which is the conversion factor found for ULIRGs at low redshifts \citep{downes98}. This results in a total molecular gas mass of M$_{\rm H2}$\,=\,(3\,$\pm$\,1)\,$\times$\,10$^{9}$ M$_{\odot}$ for the gas in the tidal debris between the two galaxies, and (2\,$\pm$\,1)\,$\times$\,10$^{9}$ M$_{\odot}$ for the gas in the outflow from the AGN host-galaxy. We note that these are likely lower limits to the molecular gas masses, as values of $\alpha_{\rm CO}$\,$\sim$\,3.6 M$_{\odot}$\,(K km\,s$^{-1}$ pc$^{2}$)$^{-1}$ are typically found in high-$z$ galaxies \citep[e.g.,][]{daddi10,genzel10}. These mass estimates for the tidal debris and outflow  are in good agreement with molecular gas masses seen in the tidal bridge in nearby merging galaxies \citep{lisenfeld02,braine03,daSilva11}, and outflows in low-$z$ ULIRGs and quasars \citep{cicone14}. The combined molecular mass of the tidal debris and outflow in the Dragonfly system represents roughly 20 percent of the total molecular gas mass detected in the combined disks of NW and SE (Paper {\sc II}). 

To derive the re-distribution rate for the molecular gas in the tidal debris that bridges the two galaxies, we assume that the molecular gas travels the 4 kpc distance between the galaxies at a typical velocity of $v$\,$\sim$\,200 km\,s$^{-1}$. This results in a mass re-distribution rate of $\dot{M}_{\rm tidal}$\,=\,170\,$\pm$\,40 M$_{\odot}$\,yr$^{-1}$ for the gas in the tidal debris and bridge. For the mass outflow rate of the NW outflow, we assume that the outflowing gas travels across the distance spanned by half a tapered beam, or $\sim$1\,kpc, with a bulk outflow velocity of \(\Delta v\,+\,\frac{1}{2}\,FWHM_{\rm CO}\,\sim\,545\) km\,s$^{-1}$ (see Sect. \ref{sec:outflow}). This results in a total outflow rate of molecular gas from the NW galaxy of $\dot{M}_{\rm outfl}$\,=\,1100\,$\pm$\,550 M$_{\odot}$\,yr$^{-1}$. 

\subsubsection{Evolution of the Dragonfly system: effects of merger, outflow, and star formation}

After having derived mass displacement rates of the molecular gas in both the tidal debris and the AGN-driven outflow, we can further investigate the relative importance of the various physical phenomena in the Dragonfly system. While the displacement of molecular gas in the Dragonfly occurs at a much faster rate in the outflow compared to the tidal debris, the relative effect that the outflow and gravitational forces have on the overall mass displacement critically depends on the lifetime of the respective events. Jet activity in radio galaxies may last for 10$^{7-8}$ years \citep{blundell01}, but is likely orders of magnitude less for (episodic) activity in young radio sources \citep[e.g.,][]{nyland20}. On the other hand, the orbital time of the system is at least 120 Myr, when considering our observed current distance of 4 kpc between the galaxies and an orbital velocity similar to the $\sim$200 km\,s$^{-1}$ rotational speed of the disks. While it is beyond the scope of this paper to determine accurate life-times of the radio source and merger event, the fact that at least as much mass of molecular gas is locked up in the slower moving tidal debris compared to the fast outflow, makes both AGN-driven and gravitational effects important for the evolution of the Dragonfly system \citep[see also][]{spilker22}. 

The tidal debris and outflow may not have the same effect on the overall gas reservoir, at least in the short term. Whereas the tidal debris may redistribute the molecular gas between the two galaxies, and thus potentially supply molecular gas to the NW galaxy that hosts the AGN (Sect.\,\ref{sec:disks}), the outflow removes molecular gas from the central region of the NW galaxy. However, in Paper\,II an estimate was made that the escape velocity at a distance of 4 kpc (i.e., the distance between the centers of the SE and NW galaxy) has a lower limit of v$_{\rm esp}$\,$\sim$\,1000 km\,s$^{-1}$, as derived from only the stellar mass of the Dragonfly system ($M_{*}$\,$\sim$\,5.8\,$\times$\,10$^{11}$ M$_{\odot}$; \citealt{debreuck10}). Therefore, it is likely that the molecular gas in both the outflow and the tidal debris will remain gravitationally bound to the Dragonfly system and will at some point be re-accreted.

In addition to molecular gas in the outflow and tidal debris, the high star formation rate of $SFR$\,$\sim$\,2000 M$_{\odot}$\,yr$^{-1}$ across the system means that molecular gas is being consumed by star formation at a rate that surpasses the mass outflow rate and tidal gas displacement rate. This means that star formation itself is as effective as the merger and outflow in removing the molecular gas \citep[see also][]{man19}. 

We therefore conclude that the gravitational effects of the merger, the AGN-driven outflows, and the vigorous star formation are all critical to the evolution of the Dragonfly system.

\subsection{Jet-disk interaction}
The 43 GHz observations conducted by the VLA detected a radio hot spot on the western edge of the SE galaxy. This hot spot appears to be aligned along the central axis of the radio jet (Fig.~\ref{fig:vla}). Based on this alignment, and our conclusion that NW hosts the radio-loud AGN, we theorize that the interaction of the southern jet with the gas and dust in the disk of the SE galaxy is causing the radio emission to brighten up at the location of the 43 GHz detection. This may happen because confinement of the radio lobe and compression of the magnetic field results in an increased particle density and enhanced synchrotron radiation \citep[e.g.,][]{gop91,barthel96,morganti11}

The gas kinematics in the Dragonfly may further support this scenario. First, Fig.\,\ref{fig:momentmaps} shows that the velocity dispersion of the gas in the SE galaxy is highest around the radio hot-spot ($\sigma_{\rm CO}$\,$\sim$\,140 km\,s$^{-1}$) and not in the center, as is the case for the NW galaxy. Such an enhanced velocity dispersion is expected in a region where energy and momentum are being exchanged between the radio source and the gas \citep{man21, mee22}.

Second, the detection of a potential blueshifted outflow (Sect. \ref{sec:outflow}) may further support the scenario of jet-brightening through the assumed geometry of the system. If the outflow is oriented towards us along the northern jet, then the  southern jet is receding. At high redshifts, the brightening of the receding jet cannot be explained by Doppler boosting, which causes intrinsic brightening of the jet only for relativistic particles moving towards us. Moreover, we only detect the radio hot-spot, which represent the working surface of the radio jet, and which is expected to advance at sub-relativistic speeds. Some Doppler boosting likely still occurs within the hot-spots, but this typically creates flux-density ratios of order a few between the hot-spots of a two-sided jet \citep{komissarov96}, while the Dragonfly shows a much higher flux-density ratio between the SE and NW hot-spot of $S_{\rm SW}$/$S_{\rm NE}$\,$\sim$\,14 at 43\,GHz. This suggests that the southern jet likely brightens because of its interaction with the SE galaxy, which enhances the radio flux of the system. The argument that the SE jet is likely receding would imply that the SE galaxy must be oriented behind the NW galaxy. 

In a forthcoming paper, we will further explore this scenario that the radio source in the Dragonfly brightens as a result of interaction with cold molecular gas, based on a comparison with other high-$z$ radio galaxies. Similar to the Dragonfly, a high flux-density ratio between hot-spots is seen in several other high-$z$ radio galaxies that show alignments between the radio source and CO-emitting gas reservoirs in the environment. However, exploration of this is beyond the scope of this paper and will be presented in Emonts et al (in prep).

If it is indeed the case that the radio flux is boosted as a result of a jet-disk interaction, then the Dragonfly Galaxy may not be as intrinsically radio bright as previously thought, making it perhaps an `imposter' radio galaxy in the high-z Universe, as we will explain in the next Section.

\begin{deluxetable}{llcc}
\tablecaption{Outflow kinematics from spectral-line fitting.}
\label{tab:outflow}
\tablehead{
\colhead{} & \colhead{} & \colhead{CO (North)} & \colhead{\HeII} 
} 
\startdata
Narrow & $v$ &  60\,$\pm$\,10	&  180\,$\pm$\,10 \\
  & $FWHM$   &	     270\,$\pm$\,30              &  990$^{\dagger}$\,$\pm$\,30  \\ 
Broad & $v$   	&
 -180\,$\pm$\,60		& -780\,$\pm$\,220 \\
 & $FWHM$ &	610\,$\pm$\,90        & 1810$^{\dagger}$\,$\pm$\,320   \\
\enddata
\tablecomments{All units are in  km\,s$^{-1}$. Errors reflect uncertainties in the fitting.\\
$^{\dagger}$ Corrected for instrumental broadening.}
\end{deluxetable}

\subsection{Archetypal radio source, or radio imposter?}
\label{sec:imposter}

The major merger between two gas-rich disk galaxies likely drives the star-formation activity in the Dragonfly, and gives the system its hyper-luminous IR appearance. The fact that the starburst IR luminosity is an order of magnitude higher than that of other powerful radio galaxies at $z$\,$\sim$\,2 could mean that either the pre-coalenscent merger is a short and active early phase in the evolution of high-$z$ radio galaxies, or the Dragonfly is not a typical high-z radio galaxy. 

The observed radio source in the Dragonfly is compact, with a total extent of only 9.5\,kpc. This could support the idea that the Dragonfly is in an early phase of radio-AGN activity. Our finding that the flux of the radio source is likely enhanced as a result of an interaction between the jet and the disk of the southern galaxy could also indicate that the Dragonfly is an `imposter' radio galaxy, with an intrinsic radio flux that is much lower than what is typical for high-z radio galaxies. At low redshifts, examples exist of radio galaxies for which the jet brightens as it encounters a companion galaxy \citep[e.g.,][]{lacy98,evans08,hota22}. The idea that the brightness of the synchrotron emission is enhanced when a young jet runs into dense clouds of gas has also been suggested as explanation for the relatively large fraction of young and compact radio sources seen among starbursting radio galaxies \citep{tadhunter11}. These starburst radio galaxies at low-$z$ likely have a rich interstellar medium as a result of a major merger. If a similar jet-cloud interaction is causing the radio source in the merger system of the Dragonfly to brighten, this could push the Dragonfly into the flux-limited regime by which powerful high-z radio galaxies are selected. 

This could also be an analogue to the nearby Seyfert galaxy PKS 1814-637, which has been classified as an `imposter' radio galaxy because the intrinsically weak radio jet interacts with its gaseous environment \citep{morganti11}. Interestingly, PKS 1814-637 is one of very few low-z radio galaxies that is classified as a genuine disk galaxy. In this respect, the fact that the radio source in the Dragonfly is hosted by a galaxy with a rotating gas disk is intriguing in comparison to this low-$z$ work. As a note of caution, gaseous disks have also been observed in low-$z$ early-type galaxies \citep[e.g.,][]{struve10,sansom19}, so further high-resolution optical or infra-red observations of the galaxies in the Dragonfly system are necessary to determine whether the AGN host-galaxy is a genuine disk galaxy. In this respect, \citet{drouart16} showed that other high-$z$ radio galaxies show star-formation histories that resemble those of spiral galaxies. Therefore, a one-on-one comparison between powerful radio galaxies in the low- and high-redshift Universe requires further investigation.

It is possible that enhancement of the radio flux, through interactions between the radio jets and a rich molecular environment, is a common way of ``creating'' a high-$z$ radio galaxy with bright synchrotron emission. This could mean that, in general, we may not observe a high-$z$ radio galaxy {\sl unless} its jets manage to interact with a surrounding dense medium. In the case of the Dragonfly system, the dense medium is the gas in the disk of a separate galaxy, but it may just as well involve gas in the interstellar or circumgalactic medium of the AGN host galaxy itself \citep[e.g.,][Emonts et al. in prep.]{steinbring14}. This scenario would still make the Dragonfly interesting in terms of its young and active starburst environment, but rather illustrative in terms of the jet-cloud mechanism that shapes bright radio sources. High-resolution observations of molecular gas and radio jets in other high-$z$ radio galaxies are essential to study the prevalence of flux-brightening of radio sources in molecular-rich environments, and whether this can tell us something profound about the evolutionary importance of the radio-loud phase and the general nature of powerful high-z radio galaxies.

\section{Conclusions}
\label{sec:conclusions}

We presented new ALMA and VLA data of CO(6-5), dust and synchrotron emission in the enigmatic Dragonfly Galaxy with a spatial resolution of $\sim$1\,kpc. We complemented this with new optical IFU spectroscopy from the Keck Cosmic Web Imager. Our main conclusions are:

\begin{itemize}
\item{The Dragonfly system is an ongoing major merger between two gas-rich galaxies with rotating disks, which likely triggered the high star-formation rates. Tidal debris from the merger is observed even in the high-$J$ line of CO(6-5).}
\item{The high velocity dispersion of CO(6-5) in the central region of NW, combined with the detection of hot-spots from the jet and counter-jet in our VLA 43\,GHz data, is consistent with the NW galaxy hosting the radio-loud AGN.}
\item{An interaction between the radio jet and the disk of the SE galaxy likely causes the jet to brighten at the hot-spot.}
\item{Molecular gas is being displaced within the Dragonfly system by both a fast moving outflow ($\dot{M}_{\rm outfl}$\,=\,1100\,$\pm$\,550 M$_{\odot}$\,yr$^{-1}$) and slower moving tidal debris ($\dot{M}_{\rm tidal}$\,=\,170\,$\pm$\,40 M$_{\odot}$\,yr$^{-1}$). In terms of total mass, both the outflow and the tidal debris contain similar amounts of molecular gas.}
\item{The gravitational effects of the merger, the AGN-driven outflow, and vigorous star formation are all important components in the evolution of the Dragonfly Galaxy.}
\item{The Keck data, which include rest-frame UV lines, identify the Dragonfly as a Type II AGN.} 
\item{Ly$\alpha$ emission of warm ionized gas is visible across 100 kpc, with the metal-lines \CIV\ and \HeII\ detected across approximately 35 kpc, confirming the presence of a rich and extended CGM previously detected in CO(1-0) in Paper {\sc I}.}
\end{itemize}

Our results suggest that the Dragonfly Galaxy may be an imposter radio galaxy, which we only observe as a powerful radio galaxy as a result of an interaction between the radio jet and the disk of the secondary galaxy. This chance interaction provides a unique snapshot of the evolutionary history of galaxies in which a major-merger event coincides with radio activity. Future work on the Dragonfly Galaxy will further investigate the nature of radio activity and feedback in this system.

More research is needed to identify how common these systems may be in the high-z Universe. Future high-resolution studies of merger systems like the Dragonfly can help to reveal the prevalence of flux-boosted radio sources in molecular-rich environments, which may have significant consequences on our interpretation of the evolutionary importance of the radio-loud phase and our understanding of the nature of powerful high-$z$ radio galaxies.

\clearpage

\begin{acknowledgments}
  Part of this work was completed as part of the National Astronomy Consortium (NAC) program at NRAO. The National Radio Astronomy Observatory is a facility of the National Science Foundation operated under cooperative agreement by Associated Universities, Inc.

  This paper makes use of the following ALMA data: ADS/JAO.ALMA$\#$2016.1.01417.S, 2013.1.00521.S. ALMA is a partnership of ESO (representing its member states), NSF (USA) and NINS (Japan), together with NRC (Canada), MOST and ASIAA (Taiwan), and KASI (Republic of Korea), in cooperation with the Republic of Chile. The Joint ALMA Observatory is operated by ESO, AUI/NRAO and NAOJ. Some of the data presented herein were obtained at the W. M. Keck Observatory, which is operated as a scientific partnership among the California Institute of Technology, the University of California and the National Aeronautics and Space Administration. The Observatory was made possible by the generous financial support of the W. M. Keck Foundation. The authors wish to recognize and acknowledge the very significant cultural role and reverence that the summit of Maunakea has always had within the indigenous Hawaiian community.  We are most fortunate to have the opportunity to conduct observations from this mountain. Based on observations with the NASA/ESA Hubble Space Telescope obtained from the Data Archive at the Space Telescope Science Institute, which is operated by the Association of Universities for Research in Astronomy, Incorporated, under NASA contract NAS5-26555.

  Support for program numbers HST-AR-16123.001-A and HST-GO-16891.002-A was provided through a grant from the STScI under National Aeronautics and Space Administration (NASA) contract NAS5-26555. DT acknowledges partial support from the National Science Foundation via award AST-2007023 to The Ohio State University. MVM acknowledges support from grant PID2021-124665NB-I00 by the Spanish Ministry of Science and Innovation (MCIN) / State Agency of Research (AEI) / 10.13039/501100011033 and by the European Regional Development Fund (ERDF) "A way of making Europe". 
\end{acknowledgments}

\vspace{5mm}
\facilities{ALMA, VLA, Keck}

\software{CASA \citep{casateam22}, KCWI data reduction pipeline \citep{neill18}  
          }





\end{document}